\documentstyle[psfig]{article}
\def\m@thcombine#1#2{%
  \setbox0=\hbox{$#1$}
  \setbox1=\hbox{$#2$}
  \ifdim\wd0>\wd1
    \setbox0=\hbox to\wd1{\hss\box0\hss}
  \else
    \setbox1=\hbox to\wd0{\hss\box1\hss}
  \fi
  \mathop{\vcenter{
    \offinterlineskip\box0\box1}}}
\def\lesim{\m@thcombine<\sim}
\def\gesim{\m@thcombine>\sim}
\def\lessgtr{\m@thcombine<>}
\def\gtrless{\m@thcombine><}

\newcommand{\ket}[1]{\left| #1 \right\rangle}
\newcommand{\omegaI}{\omega_I}

\newcommand{\eps}{\epsilon}
\newcommand{\beq}{\begin{equation}}
\newcommand{\beqa}{\begin{eqnarray}}
\newcommand{\eeq}{\end{equation}}
\newcommand{\eeqa}{\end{eqnarray}}
\newcommand{\Yb}{${}^{168}$Yb\ }

\begin{document}

\rightline{YITP-97-4, February 1997}

\vspace{5mm}

\large

\begin{center}
{\bf Level Statistics of Near-Yrast States in Rapidly Rotating Nuclei}
\end{center}

\large

\vspace{5mm}

\begin{center}
 M. Matsuo, T. D\o ssing$^{a}$, E. Vigezzi$^{b}$, and S. \AA berg$^{c}$
\end{center}

\vspace{5mm}

\large

\begin{center}
{\it
Yukawa Institute for Theoretical Physics, Kyoto University,
Kyoto 606-01, Japan \\
${}^{a}$ Niels Bohr Institute, University of Copenhagen,
Copenhagen \O, Denmark \\ 
${}^{b}$ INFN sez. Milano, University of Milan, Milan, Italy\\
${}^{c}$ Department of Mathematical Physics, Lund Institute of
Technology, Lund, Sweden
}
\end{center}

\vspace{5mm}

\normalsize

\begin{abstract}
The nearest neighbour level spacing distribution and the
$\Delta_3$ statistics of level fluctuations
associated with very high spin states ($I \gesim 30$)
in rare-earth deformed nuclei are analysed by means
of a cranked shell model. The many particle-many hole configurations
created in the rotating Nilsson potential are mixed by the  surface-delta
two-body
residual interaction. 
The levels
in the near-yrast region  
show  a Poisson-like level spacing distribution. As the intrinsic
excitation energy $U$ increases, 
the level statistics
shows a gradual transition from order to chaos, reaching at $U \gesim 2$ MeV
the 
Wigner distribution
typical of the Gaussian orthogonal ensemble of random matrices.
This transition is caused by the residual two-body interaction.
On the other hand, the level spacings between the
yrast and the first excited state show a peculiar behaviour,
displaying a Wigner-like
distribution instead of the Poisson-like distribution seen for
the other near-yrast rotational states. The lowest spacings
reflect the properties  of  the single-particle orbits
in the mean-field, and are only weakly  affected by the residual two-body
interaction.
\end{abstract}

\small
\vspace{5mm}

\normalsize

\vspace{5mm}

\section{Introduction}\label{sec:intro}

The statistical  fluctuations of the energy levels and the transition
strengths measured in 
highly excited nuclei with excitation energy above
the neutron threshold (several MeV) are 
well described by the random matrix theory 
\cite{RMT,Porter,Mehta}. For example, 
the nearest-neighbour level spacing distribution (NND) and the
spectral rigidity (or $\Delta_3$ statistics) of the neutron resonance
states follow the behaviour predicted  by the random matrix theory
for the Gaussian orthogonal ensemble
(GOE) \cite{RMT,neutronres}. 
This seems to indicate that such excited nuclei,
at least over a time scale associated 
with the observed energy interval,
are an example of a  chaotic quantal
system,  
in the sense that GOE fluctuations generally characterize 
quantum systems which are chaotic 
in the 
classical limit \cite{billiard,nuclchaos}. 

The fluctuation properties at lower excitation energy are
less well understood, although several extensive analysis of low-lying
levels as well as of near-yrast high spin levels have been
reported recently \cite{Abul,Shriner,Al,Sn,Garrett3,Garrett1,Garrett2}.
Although the low lying and low spin levels generally
show level spacing distributions  which are intermediate 
between chaos (the GOE or Wigner limit) and order 
(the Poisson limit),
one observes some systematic behaviour
with respect to the mass-number and the angular momentum
\cite{Shriner}.
In particular, it is remarkable that the NND
in heavy deformed nuclei 
is the closest to the Poisson distribution, not only for the
low-lying, low spin levels \cite{Shriner}, 
but also for the high spin rotational
levels lying near the yrast line \cite{Garrett3,Garrett1,Garrett2}.
This suggests that both the Poisson and the GOE fluctuations coexist
in rotating nuclei and that one should expect a transition from order to chaos 
with increasing intrinsic excitation energy $U$ 
(the relative excitation energy measured from the yrast line at
given spin).
In the present paper, we
examine theoretically the level statistics of high spin states
in rapidly rotating nuclei as a function of intrinsic excitation
energy $U$. In particular, we investigate in detail the
level statistics associated with the near-yrast states
which may become accessible in future experiments.
We limit ourselves to the very high spin region with $I \gesim 30$, where
static pairing is generally quenched or even vanishes,
because our model is not adequate to deal with strong pairing correlations.
Although this makes it difficult to make a direct
comparison with present  experimental
data, there are good reasons to 
expect that much more experimental information
will be available 
in the near future.

The high spin states near the yrast line in well deformed nuclei
form rotational
band structures, as evidenced
by experiments. These rotational band states are usually well described 
by  the cranked mean-field models 
\cite{Bengtsson-Frauendorf,Bengtsson-Ragnarsson,crank-rev}
in which  the collective rotation is represented by 
uniform rotation along the axis of the largest moment of inertia (axis
perpendicular to the elongated direction). The intrinsic
structure of a rotating nucleus is described in terms of the
mean-field potential adding the cranking term caused by
the uniform rotation. 
Most  observed rotational bands are based on  intrinsic configurations
with a few excited quasiparticles (or particles
and holes) defined in the cranked mean-field Hamiltonian.
However, as the intrinsic excitation energy $U$ increases at a given spin, 
intrinsic configurations
with many particles and holes ($n$p-$n$h) will show up and
become progressively dominant. Accordingly, the level density
increases significantly, reaching a value around
 $\sim 10^2$ /MeV  at intrinsic excitation energy
$U \sim 1 $ MeV above yrast line in rare earth nuclei. 
One then expects that the residual two-body interaction begins to play
an important role, mixing the $n$p-$n$h configurations,
because the size of its  matrix elements ($\sim 10$ keV) is of the same order
as the mean level spacing.
It is also to be noticed  that
the phenomenon of the rotational damping \cite{Lauritzen},
which sets in at around $U \sim 0.8$ MeV above yrast 
\cite{FAM},
is an important signature of the configuration mixing caused by
the residual two-body interaction. 
The  fluctuations of the energy levels will be sensitive to
the configuration mixing among the $n$p-$n$h configurations. 
If the configuration mixing were absent, intrinsic excitations would be
specified uniquely by the excited particles and holes. 
In such a situation, the level fluctuations
may follow the Poisson distribution.
On the other hand, once the residual
two-body interaction is switched on,
the $n$p-$n$h configurations  interact with
each other. If the residual interaction is so strong that 
many $n$p-$n$h configurations are admixed with complicated amplitudes,
one expects that the level fluctuations obey the  theory of random
matrices.
It is therefore important,
in studying the level fluctuations as a function of intrinsic
excitation energy,
to take configuration mixing explicitly into account.
We adopt a shell model approach, 
making use of a reasonable residual two-body interaction on top of a cranked 
mean field \cite{Aberg, Matsuo96}. 
Previous work with the cranking model \cite{Aberg} has
already discussed some general features of the order to chaos
transition although it used a schematic residual interaction
represented by a constant with random sign.
We have recently shown that the cranked
Nilsson model combined with the surface-delta interaction (SDI)
\cite{Mozkowski,Faessler} can reproduce the overall features of 
rotational damping found in 
experiment
\cite{Matsuo96, Matsuo93, Bracco}.
In the present paper  we adopt
the same model, studying 
the excited levels lying up to about 2 MeV above yrast 
line. We study in particular detail the states
close to yrast,
which are likely to be observed in near future experiments.

Statistical analyses of high spin levels in deformed nuclei
on the basis of the interacting boson model 
\cite{ibm}, the interacting boson fermion model \cite{IBFM}, and
the particle-rotor model \cite{Kruppa} have also been reported.
These models, however, 
take into account only  limited
degrees of freedom ($sd$ collective bosons or high-$j$ nucleons) 
of the intrinsic excitations 
in deformed rotating nuclei.

\section{Formulation}\label{sec:form}

\subsection{The model}

We start with the cranked Nilsson single-particle
Hamiltonian
\beq
	h_{crank} = h_{Nilsson} - \omega j_x  
\label{nilham}
\eeq
in order to define the single-particle basis in a rotating
deformed nucleus.
Here the quadrupole and hexadecapole deformations are
considered. We do not include the static pairing potential in
the mean-field. This may be justified for the high spin
region ($I \gesim 30$) which we are mostly concerned with, since 
the pairing gap is usually reduced, or even vanishes, 
due to the rotational perturbation
(Mottelson-Valatin effect) \cite{Garrett-pair,Shimizu,Shimizu-Oak}. 
The eigen-solutions of the cranked Nilsson single-particle Hamitonian
define an adiabatic basis as a function of the rotational frequency
$\omega$. However, since the adiabatic orbits sometimes 
accompany avoided crossings between orbits which cause abrupt change of the
basis wave functions against small change in $\omega$, 
we instead use a diabatic single-particle 
basis, which is constructed by removing small interactions causing the
repulsions at the avoided crossings. 
Putting $N$ neutrons and $Z$ protons in the diabatic
single-particle basis, shell model many-body configurations (labeled by $\mu$)
 are  generated:
\beq
\ket{\mu (I)} = \prod_{{\rm occupied}\ i \ {\rm in} \ \mu} a_i^{\dag} \ket{-}.
\label{mu}
\eeq
In Eq.(2) $a_i^{\dag}$ denotes the nucleon creation
operator for an occupied diabatic single-particle orbit $i$, which is
defined at an average rotational frequency $\omega_I$ corresponding to
the given angular momentum $I$. We include all the single-particle
orbits within an interval of 3.0 MeV below and above the
Fermi surface. 
The shell model basis $\{ \ket{\mu(I)} \}$ 
includes the  configuration in which 
the single-particle orbits up to the Fermi surface are fully
occupied, as well as all possible $n$p-$n$h configurations with respect
to the fully occupied one.

The energy of a shell model configuration $\ket{\mu(I)}$
is given, following the standard cranked Nilsson-Strutinsky prescription,
by 
\beq
E_{\mu}(I) = E_{\mu}^{Nils}(I) - E^{smooth}(I) + E^{RLD}(I)
\label{Str}
\eeq
where $E_\mu^{Nils}(I)=E'_\mu(\omega) +\omega J_{x,\mu}(\omega) $ with 
the angular momentum constraint  $J_{x,\mu}(\omega) = I$ on the rotational
frequency $\omega$. Here $E'_\mu(\omega)=\sum_{i \in \mu}
e'_i(\omega)$  and 
$J_{x,\mu}(\omega)=\sum_{i \in \mu} j_{x,i}(\omega)$ 
are the total routhian 
and the expectation value of the angular momentum $J_x$ of
the shell model basis $\mu$, respectively. 
Since we use the diabatic single-particle basis which depends only weakly
on the
rotational frequency, the energy expression can be  accurately
approximated locally by
\beq
E_{\mu}^{Nils}(I)  =  E'_{\mu}(\omegaI) + \omegaI I + 
{(I - J_{x,\mu}(\omegaI))^2 \over\ 2 J^{(2)}_{\mu}} 
\label{eng}
\eeq
referring to the average rotational frequency $\omega_I$.
Here $J^{(2)}_{\mu}$ is the dynamical moment of
inertia of the configuration.
The deviation $\left | J_{x,\mu}(\omega_I) - I \right |$ 
in the angular momentum expectation value is less than 5
at spin $I=50$ for most configurations in the present calculation.
Although the Strutinsky smoothed energy  $E^{smooth}(I)$  and the
rotating liquid drop energy $E^{RLD}(I)$ correct the absolute excitation 
energy, they
do not affect the level statistics discussed in the present paper.

We then introduce a two-body force, mixing the shell-model configurations.
We adopt the surface delta interaction
(SDI)\cite{Mozkowski}
\beq \label{eq:SDI}
v(1,2)^{angle} =  - 4\pi V_0 \sum_{LM}Y^{*}_{LM}(\theta_{t,1} \phi_{t,1})
Y_{LM}(\theta_{t,2} \phi_{t,2})
\eeq
where $(\theta_{t} \phi_{t})$ is the angle variable in the stretched
coordinates. The strength parameter $V_0$ includes the radial matrix
elements and we use the strength $V_0=27.5/A$ MeV given by
Ref.\cite{Faessler}, which is the same value used for the study of 
rotational damping in \Yb \cite{Matsuo96,Matsuo93}.
The shell model Hamiltonian is given by
\beq
H(I)_{\mu\mu'} = E_{\mu}(I) \delta_{\mu\mu'} +  
V(I)_{\mu\mu'} 
\eeq
where $V(I)_{\mu\mu'}$ denotes the matrix elements of the
residual two-body interaction of SDI. The Hamiltonian is
diagonalized to obtain
energy eigenstates
\beq
\ket{\alpha(I)} = \sum_\mu X^{\alpha}_{\mu}(I) \ket{\mu (I)} 
\eeq
which are admixtures of the basis configurations
$\{ \ket{\mu (I)}\}$ as well as their energy levels
$\{ E_\alpha(I)\}$.
The diagonalization is done separately for each $I^\pi$,
truncating the basis by including the lowest 1000 $\ket {\mu}$ basis states. 
The resulting lowest
300 states 
(covering the region up to $U \sim 2.4$ MeV) 
are rather stable against the truncation of the
basis. 
For further details of the
model, we refer to Ref.\cite{Matsuo96}. 

In the present paper, we focus on the rare-earth nuclei, and 
in particular we consider 40 nuclei in the $A = 160 -174$ region,
listed in Table 1, for which  deformed prolate shape stable up to 
very high spins is suggested by potential energy surface 
calculations \cite{PES,Werner}.
We adopt the equilibrium deformation parameters 
taken from Ref.\cite{Def-parm}, as given in Table 1,
which  are similar to those calculated in Ref.\cite{PES}.
In order to 
make a statistically meaningful analysis, we collect
the spacings  taken from  a certain spin interval 
in  all the  40 nuclei, and
we will not discuss the dependence on individual nuclei,
spins, and parities.

In the following, we use the parity  and the
signature quantum number $(\pi,\alpha)$ to classify the
energy levels of the total system. The signature $\alpha$ is related
to the total spin $I$ through the relation $I = I_0 + \alpha$
with $\alpha=0,1$ for even-$A$ system, and $\alpha=\pm1/2$ for odd-$A$.
We sometimes use the even integer spin $I_0$ and the signature
$\alpha$ in place of the ``true'' spin $I$ when we specify  spin
intervals.

\begin{table}
\begin{center}
\begin{tabular}{|c|c c|}
\hline
& $\epsilon_2$ & $\epsilon_4$ \\
\hline
$^{160,161}$Dy $^{161,162}$Ho & 0.248 & -0.016 \\
$^{162,163}$Dy $^{163,164}$Ho & 0.261 & -0.007 \\
$^{164,165}$Dy $^{165,166}$Ho & 0.267 & 0.003 \\
$^{162,163}$Er $^{163,164}$Tm & 0.245 & -0.009 \\
$^{164,165}$Er $^{165,166}$Tm & 0.258 & 0.001 \\
$^{166,167}$Er $^{167,168}$Tm & 0.267 & 0.012 \\
$^{166,167}$Yb $^{167,168}$Lu & 0.246 & 0.004 \\
$^{168,169}$Yb $^{169,170}$Lu & 0.255 & 0.014 \\
$^{170,171}$Yb $^{171,172}$Lu & 0.265 & 0.025 \\
$^{172,173}$Yb $^{173,174}$Lu & 0.269 & 0.036 \\
\hline
\end{tabular}
\caption{\label{tabdef} 
The quadrupole and hexadecapole 
deformation parameters $\epsilon_2$ and $\epsilon_4$
used in the present calculations}\end{center}
\end{table}

\subsection{Level statistics}

In order to perform the statistical analysis of the energy level fluctuations,
one must take into account the fact 
that the level density and hence the
level spacing are strongly dependent
on the intrinsic excitation energy $U$.
In this situation, it is necessary to 
separate  local level fluctuation
from the overall excitation energy dependence of the level spacings.
For that purpose, we adopt the unfolding procedure 
\cite{unfolding,billiard} in a
particular form  which follows Shriner et.al. \cite{Shriner}. 
The unfolding procedure measures the local level fluctuations 
with respect to a smooth average level density.
We assume that the average level density is represented
by the constant temperature formula
\cite{CTF}
\beq
\bar{\rho}(E)={1 \over T} \exp\left({E-E_0 \over T}\right)
\label{rhoctf}
\eeq
for each spectrum at a given $I^\pi$. 
To determine the parameters in the formula, 
we  make a fit to the staircase function which represents the
cumulative number of levels below
energy $E$,
\beq
N(E)=\int_{-\infty}^{E} \rho(E')dE' = \sum_{\alpha} \theta(E-E_\alpha) \ ,
\label{staircase}
\eeq
with a smooth function corresponding to the average
level density $\bar{\rho}(E)$,
\beq
\bar{N}(E)=\int_{E_0}^{E}\bar{\rho}(E')dE' + N_0 =\exp\left({E-E_0 \over
T}\right) -1 + N_0\ \ ,
\label{staircaseX}
\eeq
by minimizing the quantity
\beq
G(T,E_0,N_0)= \int_{E_{min}}^{E_{max}} \left( N(E) -\bar{N}(E) \right)^2 dE
\eeq
with respect to the parameters $T, E_0$ and $N_0$ for each spectrum
at a given $I^{\pi}$.
Here the energy boundaries $E_{min}$ and $E_{max}$ are the energies
of the lowest and the 300-th levels in each spectrum.
When we discuss the level statistics for the lowest 20 levels,
however, we obtain a better fit including only the lowest 30 levels. 
The unfolded spectra $\{ x_\alpha ; \alpha = 1,2,... \}$ are then derived
for each $I^{\pi}$ by the transformation 
\beq
x_\alpha  = \bar{N}(E_\alpha). 
\eeq
The unfolded spectra have a constant average density
$\bar{\rho}_x(x) =1$ provided that the constant temperature formula
fits well the average level density.   

In order to analyze the level fluctuations, we calculate the
nearest neighbour level spacing distribution (NND) which is also often
used in the experimental analysis. 
We calculate the distribution
$P(s)$  for the unfolded
spectra, where $ s= x_{\alpha+1} -x_\alpha$
is the spacing between the neighbouring levels with same $I^\pi$. 
By the unfolding procedure, the spacings
are normalized as $\langle s \rangle = 1$.
The distribution is represented as a histogram.

The NND is calculated for various ensembles of 
level spacings 
which are taken from different intervals in excitation energy.
The obtained distribution is fitted with the Brody distribution \cite{RMT}
\beq
P_w (s) = (1+w)\alpha s^w\exp(-\alpha s^{1+w}),\ \ 
\alpha = \left(\Gamma\left({ 2+w \over 1+w}\right)
\right)^{1+w} \ \ ,
\label{Brody}
\eeq
parametrized by the Brody parameter $w$.
This family of distributions is convenient because 
the Brody parameter $w=1$ produces 
the Wigner distribution while the value $w=0$ 
corresponds to the Poisson distribution. (Note that 
the theory of GOE random matrices leads to $w=0.953$
\cite{RMT} , which 
is not distinguishable from the Wigner limit in the present analysis).
The value of $w$ is determined minimizing the quantity
\beq
S(w) = \sum_i \left( {P(i) - P_{w}(i)} \over {\sigma(i)} \right)^2 \ ,
\eeq
where $P(i)=N(i)/N$ is the probability 
in the $i$-th bin $[s_i, s_i+\Delta s]$ of the calculated NND
($N(i)$ being the number of spacings in bin $i$ 
out of the total spacings of $N$). $P_w(i)=\int_{s_i}^{s_i+\Delta s}
P_w(s')ds'$ is the corresponding
probability in the Brody distribution.
The statistical error is estimated as $\sigma(i) = \sqrt{N(i)}/N$
for $N(i) > 0$ and $\sigma(i) = 1.15/N$ for $N(i) = 0$ 
by assuming the multinomial distributions. 

We also calculate the ensemble average of 
the $\Delta_3$ statistics \cite{delta3,RMT}
\beq
\bar{\Delta}_3(L)= \left\langle{ 1\over L} \mathop{\min}_{A,B} 
\int_{x}^{x+L} \left[N_x(x') -Ax'-B\right]^2 dx'\right\rangle
\eeq
or the spectral rigidity. Here 
$N_x(x)=\sum_\alpha\theta(x-x_\alpha)$ 
is the staircase function for the unfolded spectra and the
average $\langle ... \rangle$ is calculated over spectra
in a given ensemble and intervals $[x,x+L]$, $[x+L/2,x+3L/2]$,
... in a spectrum \cite{billiard}.
For the Poisson distribution of 
levels, 
\beq
\bar{\Delta}_{3,{\rm Poisson}}(L)=L/15 \ \ ,
\eeq
while for the GOE  distribution,
\beq
\bar{\Delta}_{3,{\rm GOE}}(L) \approx {1 \over \pi^2} (\ln L - 0.0687) 
\ \ .
\eeq

\section{Results and Discussion}\label{sec:results}

\subsection{Order to chaos transition}\label{sec:nnls}

We first discuss how the level statistics depends on the 
intrinsic excitation  energy $U$, aiming at
extracting the overall dependence on $U$
in a wide interval ranging from $U=0$ (at yrast line) to
$U \sim 2$ MeV.  For that purpose, we calculate the NND
and 
$\Delta_3$ for the lowest 300 levels in each spectrum, 
grouping them in bins of levels.
The intrinsic excitation energy of the binned
levels approximately
covers the region up to $U \sim 2.4$ MeV.

\begin{figure}
\centerline{\psfig{figure=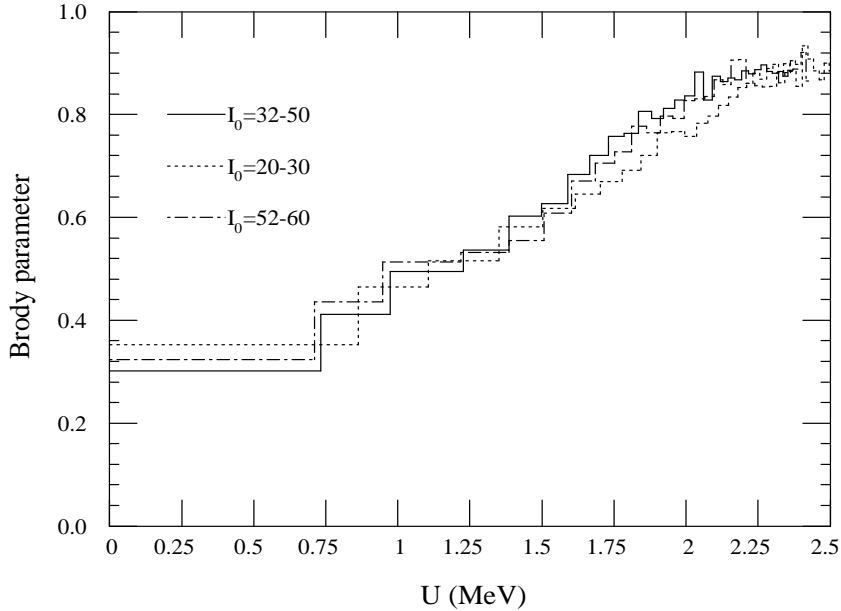,height=8cm,angle=-90}}
\caption{\label{fig1} 
The Brody parameter extracted from the NND
for energy bins containing the first to 5-th, 6-th to 10-th,
11-th to 20-th, 21-st to 30th, and 31-st to 40-th levels, ...
291-st to 300-th of each spectrum. The result is plotted as a function of
the  intrinsic excitation energy covered by the bins.
The solid, dotted, and dot-dashed lines correspond to
different spin intervals $I_0=32-50,
20-30, 52-60$ respectively. 
}
\end{figure}

\begin{figure}
\centerline{\psfig{figure=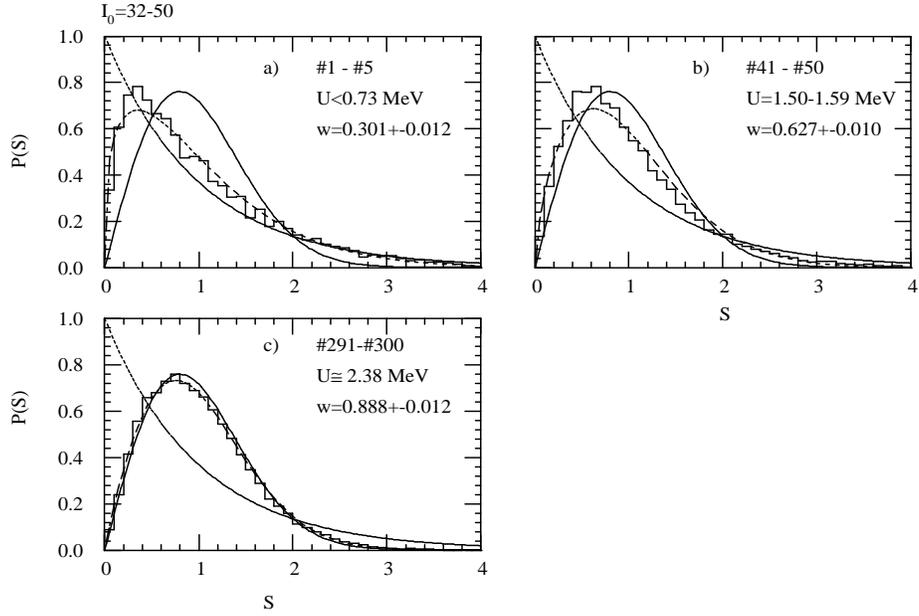,height=8cm,angle=-90}}
\caption{\label{fig2} 
The NND
for energy bins containing the first to 5-th, 41-st to 50-th, and
291-st to 300 th levels of each spectrum within spin interval $I_0=32-50$. 
}
\end{figure}

The calculated Brody parameter for the NND
for the spin interval $I_0 = 32-50$ is depicted in Fig.\ref{fig1}.
The Brody parameter increases monotonically
with increasing intrinsic excitation energy $U$.
The NND
for the lowest
bin (first to 5-th levels at each $I^\pi$),
has the Brody parameter $w=0.301\pm0.012$. The corresponding
NND shown in Fig.\ref{fig2}(a)
is much closer to the Poisson than to the Wigner distribution,
although one can also notice a small deviation from the Poisson  distribution.
For the levels from 10-th to 40-th, the Brody
parameter is about 0.5, which is   midway between the
Poisson and the Wigner distributions, as can also be seen from the
NND plotted in Fig.\ref{fig2}(b). 
As the intrinsic excitation energy
increases further, the NND approaches  the
Wigner limit; the bin including the levels from
the 291-st to 300-th  has 
$w=0.888 \pm 0.012$, which is close to
the Wigner limit $w=1.0$ or GOE limit 0.953 (See also the NND
shown in Fig.\ref{fig2}(c)).
These results indicate that the transition from order
(Poisson fluctuation of the levels) to  chaos 
(Wigner  and GOE fluctuations) takes place gradually increasing
the intrinsic excitation energy, until the chaotic limit is nearly
achieved 
at around $U \sim 2$ MeV above yrast line. 
This dependence on excitation energy confirms the results of a previous
analysis \cite{Matsuo93} performed with the same model,
but without transforming 
the energy from the rotating frame to the laboratory frame
(the last two terms in Eq.(\ref{eng}) were  neglected).

\begin{figure}
\centerline{\psfig{figure=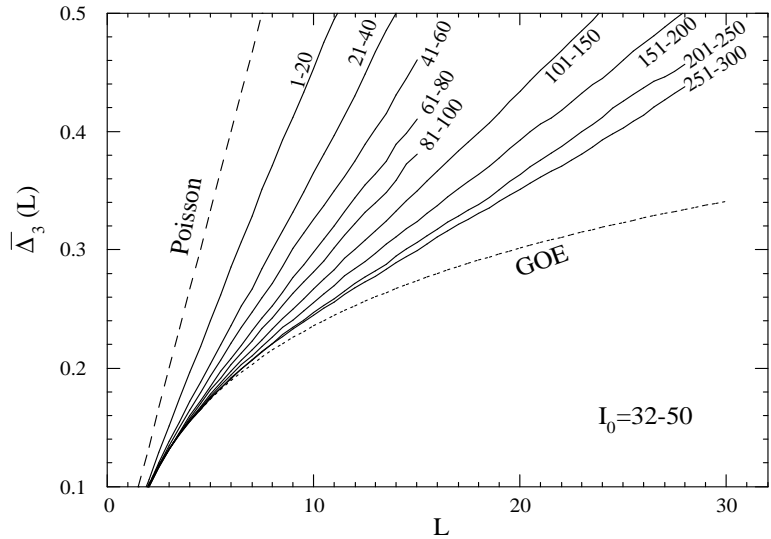,height=7cm,angle=-90}}
\caption{\label{fig3} 
The spectral rigidity $\bar{\Delta}_3(L)$ calculated 
for different energy bins containing 20 or 50  levels 
in each spectrum with fixed $I^\pi$ within the spin interval
$I_0=32-50$, plotted with solid lines. 
}
\end{figure}

It is interesting to note the implications of the results from the 
$\Delta_3$-statistics (Fig.\ref{fig3}). 
The GOE limit is obtained only for 
$L$ values up to some value $L_{\rm max}$, 
and it is found that $L_{\rm max}$ 
increases with increasing excitation energy.  For the bin at 
the highest studied excitation energy (\#251 --- \#300,  $U \sim 2.3$ MeV),
we find that $L_{max}\sim 6$. 
This implies that an energy eigenstate in this interval 
follows the GOE correlation  only with approximately the ten closest 
lying states. 
Thus, the GOE-behaviour seen in the NND for the energy levels in this 
interval (Fig.\ref{fig2}(c))  indicates 
a chaotic behaviour of local nature. 
The NND and $\Delta_3$ carry different types of information for
short-range and long-range correlations, as discussed in \cite{Pe95}.
Since the spreading width $\Gamma_\mu$ of
$n$p-$n$h shell model basis states  is finite,
$L_{\rm max}$ could be related to $\rho \Gamma_\mu$ ($\rho$ being 
the level density) \cite{Aberg,Pe95} although
an estimate based on $L_{\rm max} \approx 2.5 \rho \Gamma_\mu$ 
which is found in a random matrix model \cite{Pe95} gives a much larger
value than the calculated $L_{\rm max} \sim 6$.
Non generic behaviours of $\Delta_3$ have also been
discussed in connection with the shortest periodic orbits \cite{Berry} 
and the Lyapunov exponent \cite{Arve} in semiclassical analysis, whose
relation to the present model is not clear yet.

In Fig.\ref{fig1}, we also show the Brody parameters extracted
from ensembles of binned levels taken from lower
and higher spin intervals with $I_0 =20-30$ and $I_0 = 52 - 60$.
No significant spin dependence is observed. This is in contrast with
the interacting boson  fermion  model \cite{IBFM}
and the particle-rotor model \cite{Kruppa}, 
which predict a spin dependence  caused 
by the alignment of the high-$j$ orbitals.
Note that, besides the high-$j$ orbitals,  
we include all the other single-particle orbits near the Fermi
surface, which do not necessarily align in the considered spin 
interval.

It is interesting to compare our results with the 
previous theoretical analysis by \AA berg \cite{Aberg}, who used 
essentially the same model except for the matrix elements
of the two-body interaction,
which were schematically approximated by a constant
with random sign. When the mean square root value 
of the matrix elements was 
15 keV, the $\Delta_3$ statistics reached the GOE limit
in the excitation energy range $U=1.5 - 2.0$ MeV in \Yb, a value
lower than in the present model.
The difference can
be traced back to the statistical properties of the
two-body matrix elements. We find that the statistical distribution 
of the off-diagonal matrix elements
of the SDI force  follows a distribution strongly
peaked at zero matrix element compared with the Gaussian distribution
\cite{Matsuo96,Matsuo93},
indicating a selectivity in the two-body matrix elements related to
the intrinsic nature of the SDI and of the cranked Nilsson single-particle
orbits. Because of this selectivity, the onset of chaos in
our model takes place at higher excitation energy,
although the average mean square root of the off-diagonal matrix
elements is about 19 keV in the present calculation.

\subsection{Level statistics in the near-yrast region}\label{sec:near}

The bin including the lowest 5 levels for each $I^\pi$ covers
the interval up to $U \sim 0.7$ MeV above the yrast line. 
The calculated levels
in this energy region  mostly form
rotational band structures connected by 
strong stretched E2 transitions \cite{Matsuo96}. 
These levels are probably those which will be resolved in experiments 
in the near future, while it will be much harder to
resolve excited levels lying in the region
of rotational damping  ($U \gesim 0.8$ MeV).
In fact, up to around 10-20 rotational bands are observed
in a few rare-earth nuclei \cite{Fitz168Yb,Nord164Yb} (although
only a few rotational bands are identified at the
highest spin $I \sim 40$).
In this subsection, we discuss
in detail the level statistics associated with these
near-yrast states.  

For this  purpose, we introduce a strict ordering of 
the spacings according to excitation energy
above yrast. The strict ordering $N$
encompasses four spectra
having different parity $\pi$ and signature $\alpha$ 
for a given reference spin $I_0$ (even integer), that is
those spectra with 
$I^\pi= I_0^\pm, (I_0+1)^\pm$ for even-$A$ systems, and
those with $I^\pi= (I_0\pm1/2)^\pm$ for odd-$A$ systems.
More precisely, we first 
define  
a reference energy  $E_{ref}(I)$ by an envelope of the
yrast levels, i.e, the 
$E_{ref}(I) = \min \left\{E_{lowest}(I), {E_{lowest}(I+1)+ 
E_{lowest}(I-1) \over 2} \right\}$ 
in order to compare the four spectra. 
We then assign the label $N$ to the levels in the four spectra, 
counting from the lowest
according to the excitation energy $E(I) - E_{ref}(I)$ measured
from the reference. By this definition the $N=1$ level
represents the ``strictly yrast'' level in the sense that
it refers to only one among the four lowest levels  defined
separately for the four parity and signature quantum numbers,
while the other three
are treated as excited levels ($N>1$) with respect to the
strict yrast level.
Collecting the spacings from the $N$-th rotational
band (they are the spacings between the $N$-th level and the
next excited level with the same $I^\pi$) from the
spin range $I_0 = 32-50$ in the 40 rare-earth nuclei, 
we calculate the NND 
and extract the Brody parameter for each $N$.
We also made the same analysis for spin intervals,
$I_0=20-30$ and  $I_0=52-60$ in order to study a possible
spin-dependence,
although the present model may not be very
realistic for the lower spin interval 
($I_0=20-30$) because of the problem of the
pairing correlation.

\begin{figure}
\centerline{\psfig{figure=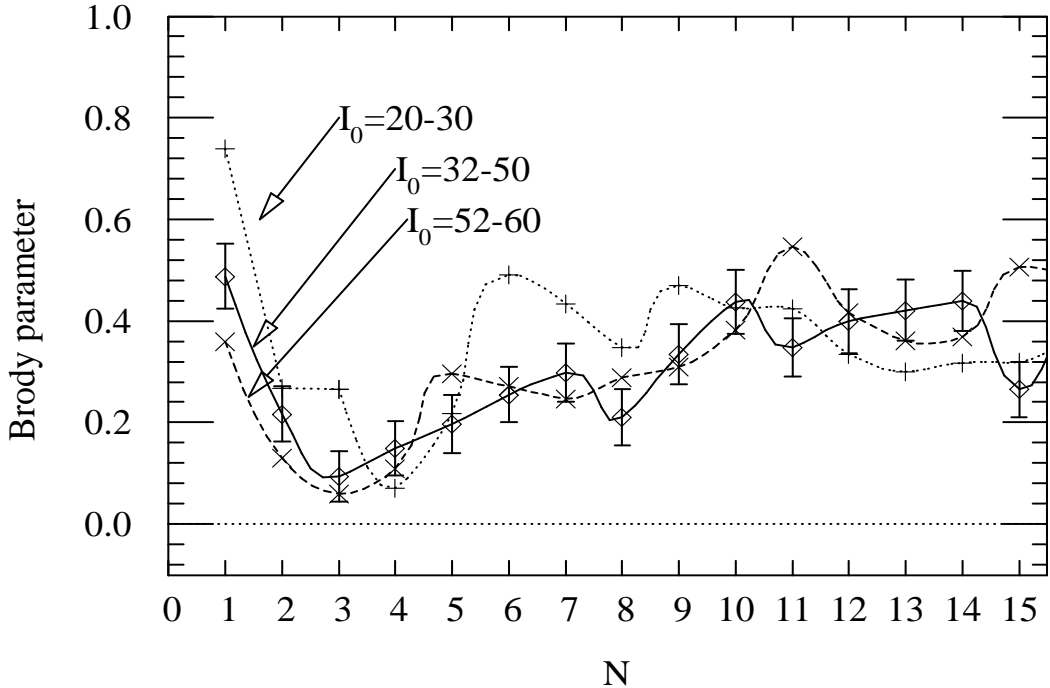,height=9cm,angle=-90}}
\caption{\label{fig4} 
The Brody parameter extracted from the NND 
associated with the lowest 15 near-yrast states.
See text for the definition of the strict ordering $N$ of
the states. Different symbols represent spin intervals
$I_0=32-50$, $I_0=20-30$, and $I_0=52-60$. 
}
\end{figure}

\begin{figure}
\centerline{\psfig{figure=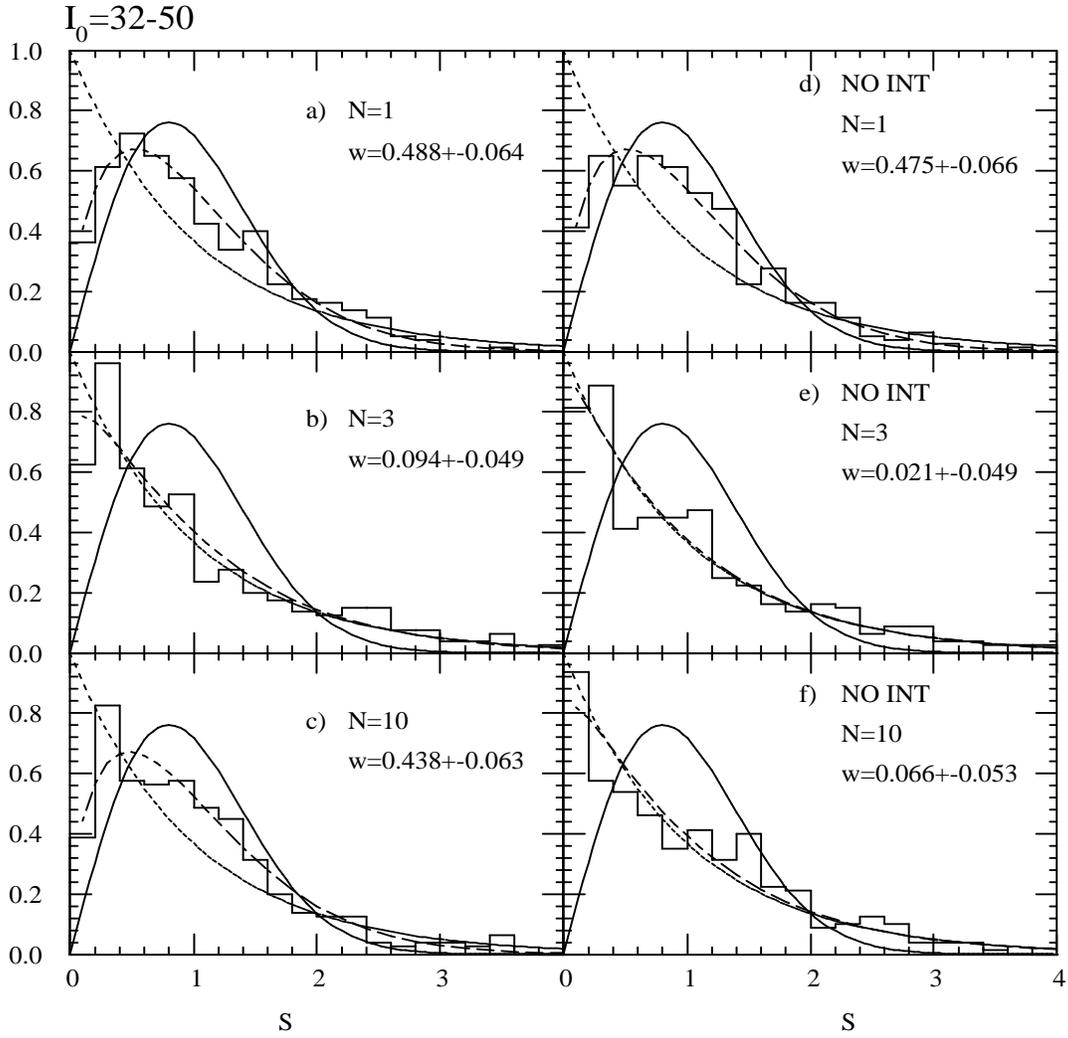,width=14cm,angle=-90}}
\caption{\label{fig5} 
The NND associated with  the lowest, third, and
the tenth
(strict order $N$=1,3,10) levels 
within the spin interval $I_0=32-50$ for (a), (b), and (c),
respectively.
(d,e,f) the same as (a,b,c) 
except that the residual SDI interaction is neglected.
}
\end{figure}

The extracted Brody parameter
is plotted in Fig. \ref{fig4}
as a function of the strict level order $N$, for the lowest 15 levels.
The spacings analysed for Fig. \ref{fig4} belong mostly within
the first energy bin 
adopted
in the previous subsection, which included the lowest 5 levels
at each $I^\pi$. The Brody parameters plotted in Fig.
\ref{fig4}
in average agree with the value 
$w \sim 0.3$ for the
lowest bin in Fig.\ref{fig1}. It is   seen in 
Fig.\ref{fig4} that the Brody parameter gradually 
decreases as the levels become closer to the yrast line, except for $N$=1.
For the lowest few levels, 
the Brody parameter is about 0.1-0.2, which is close to the Poisson 
limit (The corresponding NND's are shown in Fig.\ref{fig5}(b,c)
for the third and tenth lowest levels $N=3$, and 10).
This indicates that the excitation energy dependence
shown in Fig.\ref{fig1} continues down to the lowest 
few states near the yrast line.
However, a remarkable 
deviation from the overall excitation energy dependence
is clearly noticed  for the lowest point at $N=1$, i.e.,
for the spacings between the yrast band and the next
excited band with same $I^\pi$, for which $w \approx 0.4-0.7$. 
The NND for
$N=1 $ is shown in Fig.\ref{fig5}(a).

It is also seen that the spin dependence is not strong while
at $N=1$ the Brody parameter at lower spins shows
a  more significant
deviation from the Poisson limit, becoming close to the Wigner
limit.

\subsection{Level spacing statistics at yrast}\label{sec:yr}

\begin{figure}
\centerline{\psfig{figure=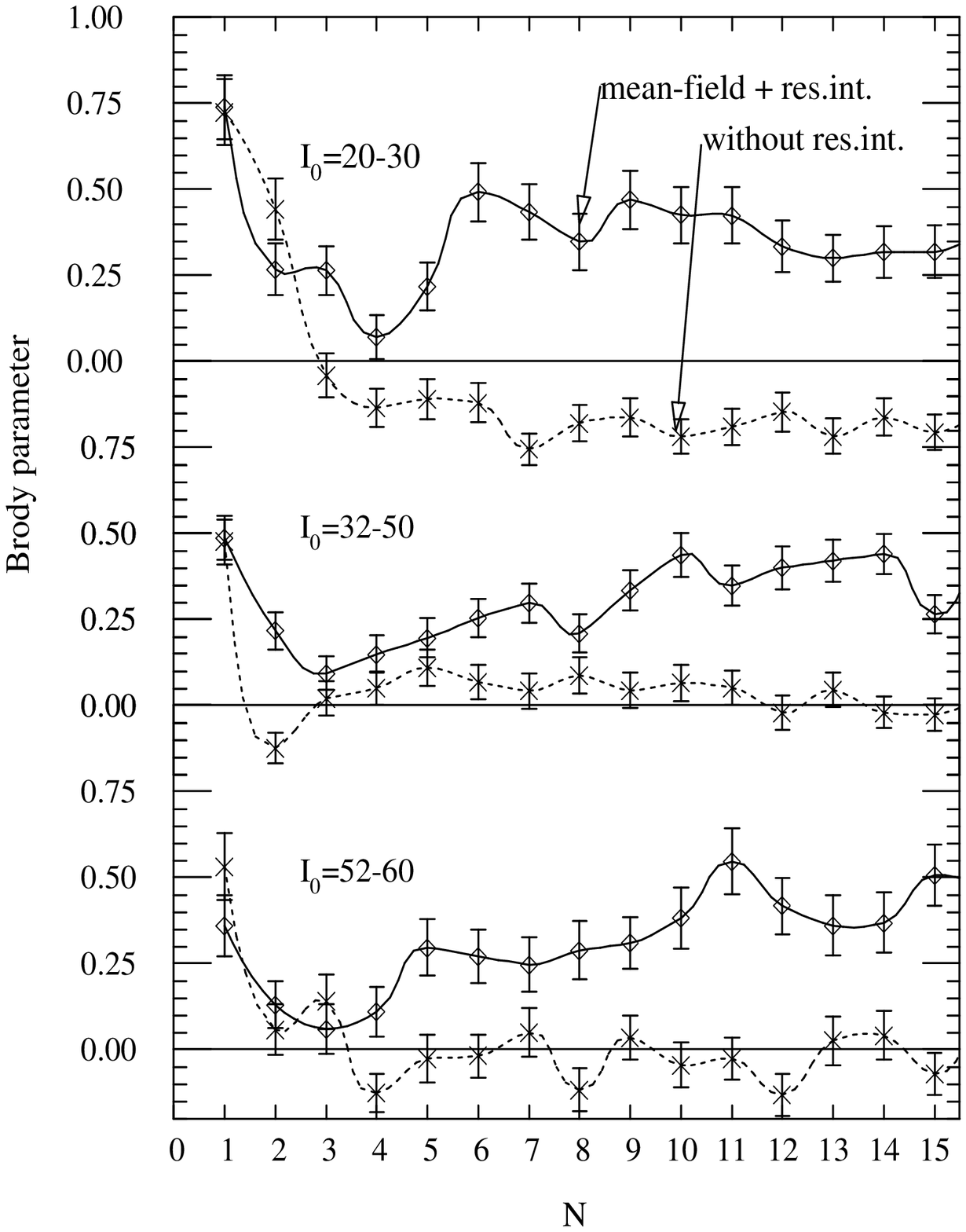,height=12cm}}
\caption{\label{fig6} 
The Brody parameter extracted from the NND 
associated with the lowest 15 near-yrast states in the strict order
obtained
from the mean-field calculation 
without the residual interaction
for different spin intervals 
 $I_0=20-30$,  $I_0=32-50$ and $I_0=52-60$ 
(points joined by dashed curves).
It is compared with the result with the residual interaction 
(ones joined by solid curves. See also Fig.\ref{fig4}).
}
\end{figure}

In order to study the origin of the  deviation of the first 
spacings from the Poisson distribution,
we perform calculations neglecting 
the residual two-body
interaction in which 
all the states become pure many-particle many-hole 
mean-field configurations.
The NND and the extracted Brody parameter are
compared in Figs.\ref{fig5} and \ref{fig6}
with those obtained by inclusion of the residual interaction.

The Brody parameter of the lowest  spacings ($N=1$) 
is essentially  unaffected
by the inclusion of the residual interaction.  
This indicates that the deviation
from the Poisson limit for $N=1$ does not originate from the
residual interaction, but from the mean field.
On the other hand,  Figures \ref{fig5}
and \ref{fig6} also show  that for the higher spacings
above $N \sim 3$, the Brody parameter of the pure mean field calculation 
converges to the Poisson limit
$w=0$ except for $I_0 = 20-30$
\footnote{For the rotational frequencies
corresponding to the spin $I_0=20-30$, many of the cranked-Nilsson
routhian orbits show very little signature splitting. This causes
frequent near degeneracy in the $n$p-$n$h configurations, leading
to the enhancement of the NND for small spacings and 
producing 
negative values  of the Brody parameter for $N \gesim 3$
in $I_0=20-30$ case in Fig.\ref{fig6}.}
while the Brody parameter and the NND calculated
with the residual two-body interaction deviates from
the Poisson limit. 
This explicitly indicates that , contrary to the case of $N=1$, 
the deviation from the
Poisson limit at $N \gesim 2$ arises from the residual two-body
interaction.

\begin{figure}
\centerline{\psfig{figure=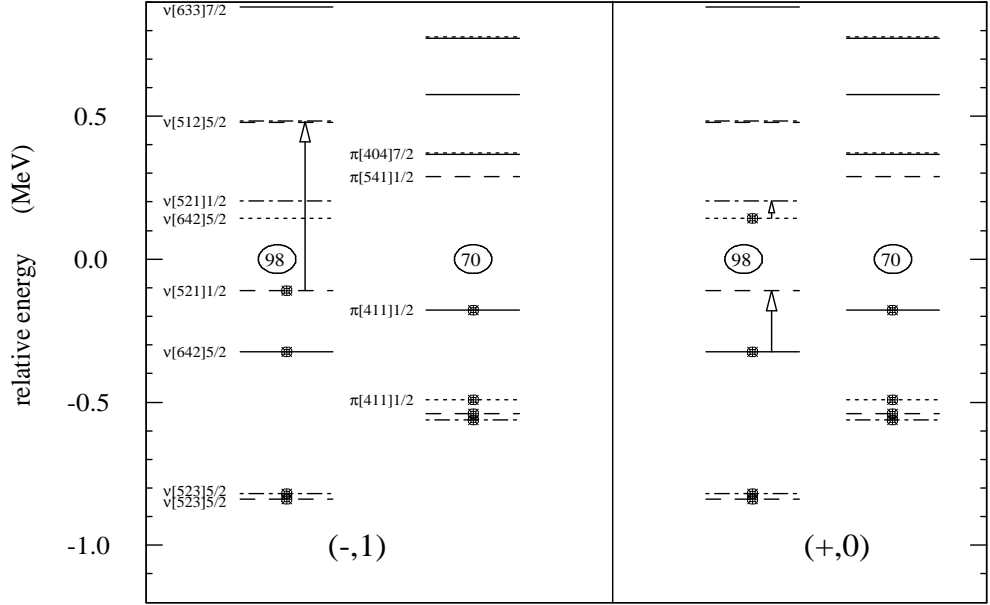,height=8cm,angle=-90}}
\caption{\label{fig7} 
Main mean-field configurations in the lowest states
with  $(\pi,\alpha)=(-,1)$ and
$(+,0)$ in \Yb at $\omega=0.319$ corresponding
to $I=30,31$. The blobs represent the
occupied orbitals in the main  configuration
of the lowest state for
each $(\pi,\alpha)$. At this spin, the yrast state belongs
to $(-,1)$.
The arrows represent  excitations
involved in the second lowest state in each $(\pi,\alpha)$.
The solid,dotted, dashed and dot-dashed lines represent
the cranked Nilsson single-particle orbits with
$(\pi,\alpha)=(+,1/2),(+,-1/2),(-,1/2)$, and $(-,-1/2)$
respectively.
}
\end{figure}

We look for the origin of the special feature of the very yrast 
$N=1$ spacings in connection with 
the single-particle level structure in the cranked Nilsson 
mean-field. To this end, we first remark that
for the levels near the yrast line, the configurations are only weakly
mixed by the residual two-body force and most of
them have essentially independent-particle configurations. In particular, 
the excited states near the yrast lines often have 1p1h configuration with
respect to the yrast configuration. This means that
the relatively large value of the Brody parameter extracted
for the yrast levels may not be directly related to the mixture
caused by the residual interaction. Furthermore, when 
the spin is not very large, the angular momentum alignments of 
intrinsic excitation is relatively small 
compared to the level spacings between the states in the
yrast band and the next excited states with the same quantum number;
it is found that the last term in Eq.(\ref{eng}) representing 
the alignment effect on the energy is at least a factor $\gesim 2$ smaller
than the average level spacing
$D \sim 350$ keV associated with the $N=1$ yrast rotational band.
Under these conditions, the relative energy of the first
excited states having the same $I^\pi$ measured from the
yrast states 
can be approximated by the 1p1h excitation energy in the single-particle
routhian
spectrum. 
Namely, $E_{next}(I^\pi) - E_{yrast}(I^\pi) \sim 
e'_{p,\alpha\pi}(\omega_I) - e'_{h,\alpha\pi}(\omega_I)$ 
where $e'_{p,h}$ are the 
single-particle routhian 
of the involved particle and hole. 
The particle and hole orbits necessarily have the same
quantum numbers (signature $\alpha$ and parity $\pi$).
On the other hand, 
spacings between excited states (yrare levels) do not keep such relation to
the single particle spacings.
This is illustrated in Fig.\ref{fig7} which shows examples of the
main mean field components of the lowest excitation within
each parity-signature set of states. It is clearly seen how the 
excitation from the yrast state in this typical case will 
proceed by changing the orbit
of one particle, keeping its $(\pi,\alpha)$ (as shown in the left
panel of Fig.\ref{fig7}). 
For yrare
levels, excitations relative to the Fermi surface are already
present, and the lowest excitation starting
from an yrare state most often will proceed by a 2p-2h excitation
which connect orbitals with different $(\pi,\alpha)$, as shown
in the right panel.

In the previous section, we find that the
the special feature seen for the yrast bands
becomes more prominent as the spin decreases.
This feature is consistent with the above interpretation.
In addition, we find that 
there exists an odd-even effect for $N \lesim 10$ for the low spins
$I_0=20-30$; the level statistics for 
odd-odd nuclei is very close to the Poisson distribution while
even-even or odd-$A$ nuclei shows significant deviation from the
Poisson distribution for $N=1$. This also indicates that the
level spacings associated with the very yrast states reflects
the single-particle level spacings since the odd-even effects
can arise from the fact that many cranked-Nilsson routhian
orbits retains two-fold degeneracy (signature splitting is small)
at low rotational frequency.

\begin{figure}
\begin{minipage}[t]{6cm}
\psfig{figure=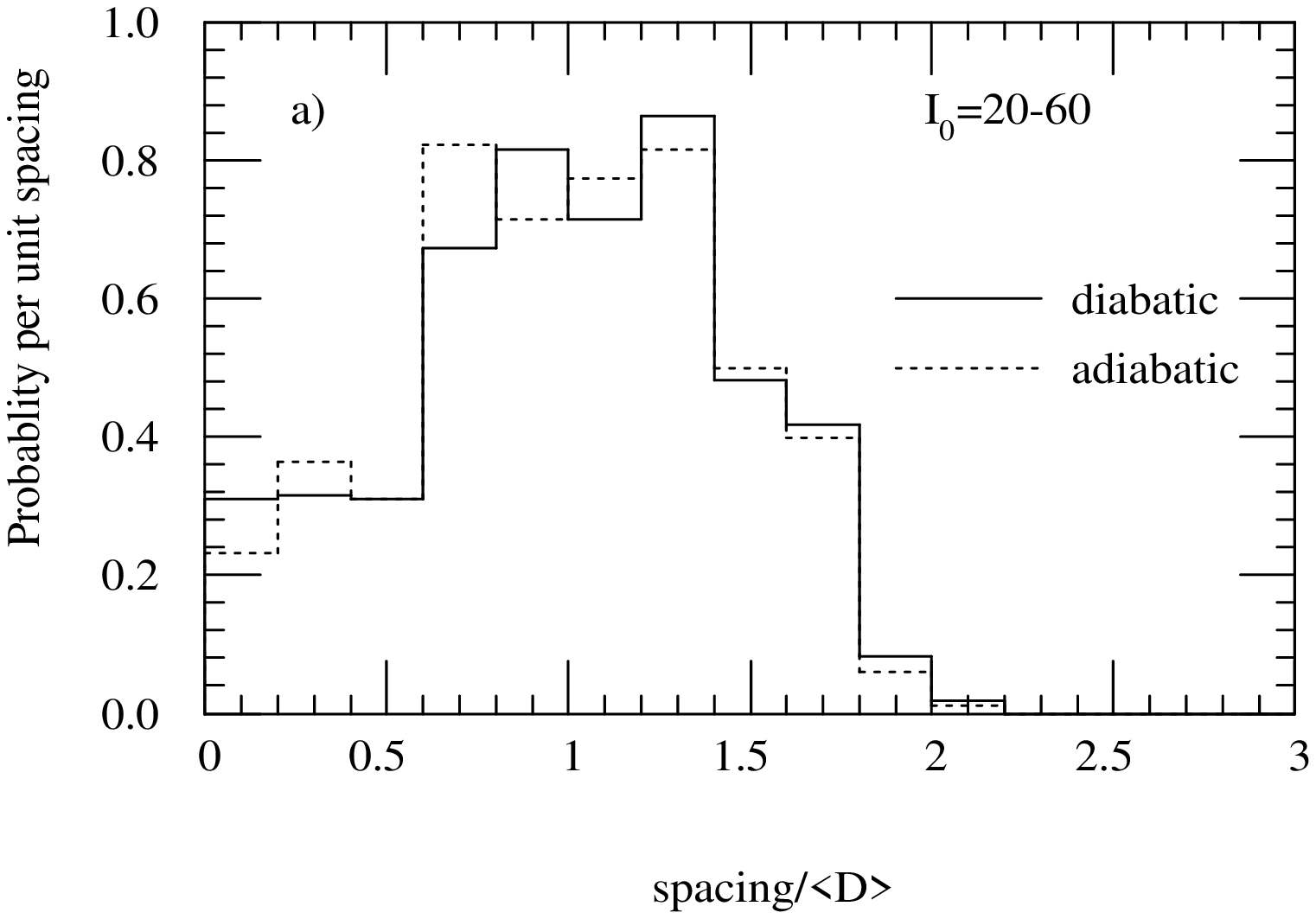,width=7cm,angle=-90}
\end{minipage}
\hskip 1cm
\begin{minipage}[t]{6cm}
\psfig{figure=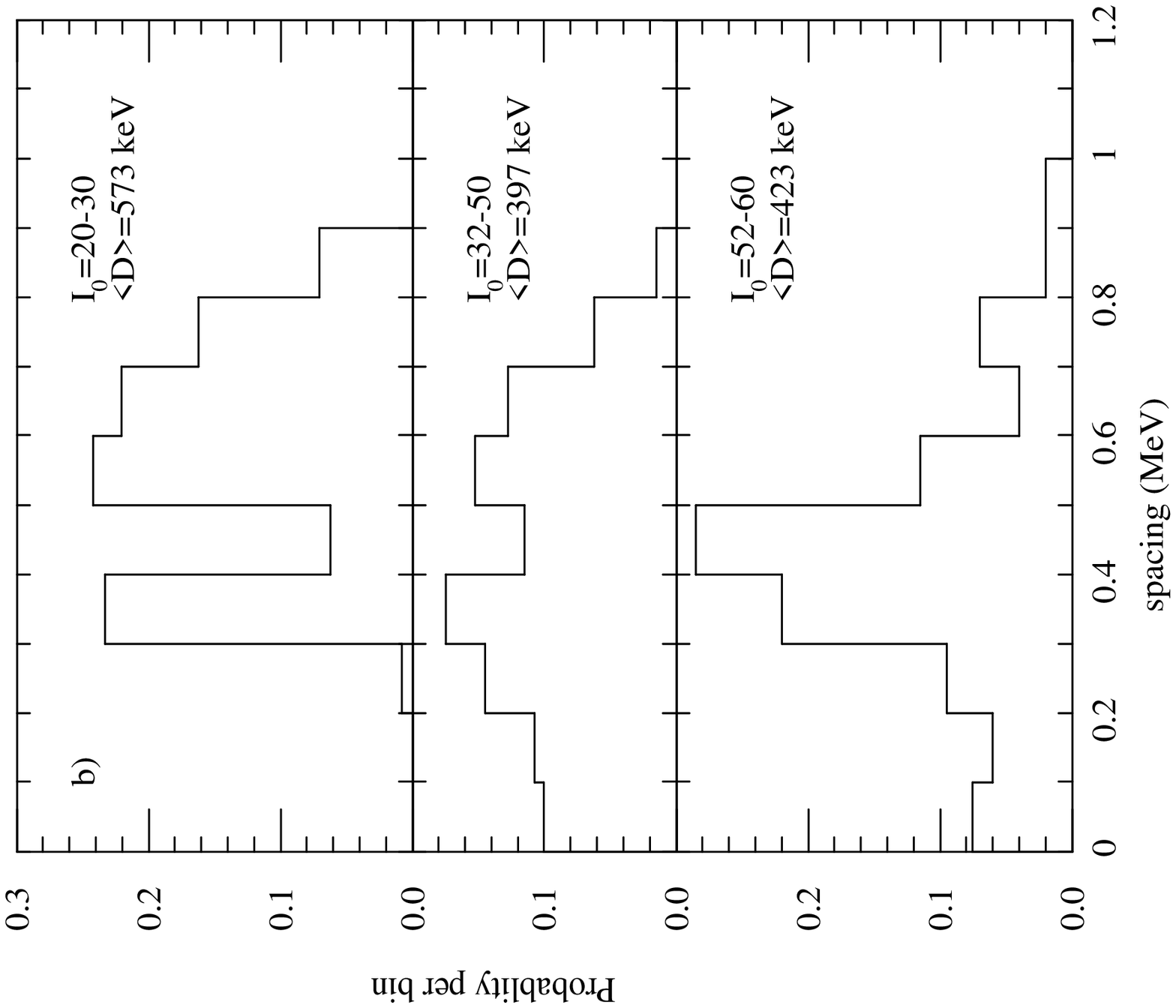,width=6cm,angle=-90}
\end{minipage}
\caption{\label{fig8} 
(a) The distribution of the spacings of
the cranked Nilsson single-particle orbits for spin interval
$I_0=20-60$. The sampling is described in the text.
The dotted line represents the distribution which is obtained
when the adiabatic basis is adopted. There is no significant
difference between the diabatic and adiabatic basis.
(b) The same as (a), but the histogram bins are defined with the
spacing itself instead of the normalized spacing, and also the spin interval
is subdivided into  $I_0=20-30$,  $I_0=32-50$ and $I_0=52-60$. 
}
\end{figure}

These considerations lead us to investigate
the spacing distribution
of the {\it single-particle levels}
in the cranked Nilsson potential.
Figure \ref{fig8} shows distribution of the level spacings
$e'_{p,\alpha\pi}(\omega_I) - e'_{h,\alpha\pi}(\omega_I)$ 
between the hole and particle orbitals having the same parity and signature
quantum number which correspond to the lowest 1p1h excitations
for all 40 nuclei and rotational frequencies
corresponding to spin $I_0=30,32,...50$. We have not applied the unfolding
procedure since the relevant single-particle orbits
lie only in a limited region around the neutron and proton 
 Fermi surfaces ($N =94-105, Z=66-71$) of the cranked 
Nilsson spectrum.
It is  noticed in Fig.\ref{fig8} 
that the spacing distribution is concentrated
around the average spacing ($\langle D \rangle \sim 400$ keV) and 
there are few spacings smaller than 200 keV, indicating that
degeneracy among orbits with the same quantum numbers happens only 
rarely. This  in fact comes from the nature of the 
cranked Nilsson spectrum (which is believed to be
valid also for other models such as Woods-Saxon  potential).
One of the relevant properties is that a large part of the deformed mean field 
is of harmonic oscillator type, and that the quantum spectrum
of the oscillator  shows strong level repulsion while the
corresponding classical motion is integrable.
With deformation not very large ($\eps \lesim 0.3$) the
single-particle orbits around the Fermi surface  belong to 
a single major oscillator shell, provided that the parity and the kind
of particle is fixed. Taking the neutron spectrum as an example,
the negative parity orbits are dominated by those with
the total oscillator quanta $N_{osc}=5$. Because of the mean-field deformation,
the orbits having different $n_3$ (oscillator quanta along the deformation
axis) are then splitted in energy, and this makes 
degeneracy among orbits with different $n_3$ asymptotic number
rare. Furthermore, 
the $l^2$ and $ls$ terms of the mean field cause splittings 
among the orbits having the same $n_3$. 
The positive parity neutron orbits near the Fermi surface are  $i_{13/2}$
orbits, and because of the deformation splitting, the $i_{13/2}$
orbits with fixed signature are
placed with finite intervals at any rotational frequency.
Therefore  degeneracy among the $i_{13/2}$ orbits never happens.
An additional mechanism arises from the fact that
the  nuclear mean-field favours an equilibrium deformation at which the shell
energy lowers, implying that degeneracy of single-particle
orbits at the Fermi surface is unfavoured. 
All these mechanisms prefer 
to the Wigner-like distribution in the
single-particle spacing at the Fermi surface. 
Consequently, there exist only few cases of  small spacings
as seen in Fig.\ref{fig8}, especially for the low spin region 
$I_0 < 30$. At higher spins (i.e., at high rotational frequency),
some specific orbits with large rotational alignment, e.g, proton orbits
stemming from $h_{9/2}$ and  $i_{13/2}$ intrudes in the Fermi
surface region around $I\sim 40-50$ 
and cross sharply with other orbits. 
Small spacings associated with these highly
aligned orbits are present  in the distributions shown in
Fig.\ref{fig8}(b) for  $I_0 = 32-50$ (and slightly also for $I_0=52-60$),
but this does not enhance very much the probability of
small spacings and keep the distribution Wigner-like.

Consequently, 
the intrinsic nature of the cranked 
single-particle spectrum 
affects specifically
the level spacing distribution associated with the yrast band
at $N=1$.
Figure \ref{fig6} indicates   that this
remains to some extent even  for the very high spin region with
$I=30-60$. By the same token, the present analysis suggests 
that the singular behaviour
of the lowest spacing could be stronger at lower spins. It should
be remarked however that the present model is not very accurate
for describing the near-yrast rotational bands  at lower spins since
the pairing correlation which is important at low spins 
is not well taken into account. 
Thus, the results of the present analysis cannot readily be 
compared to experiments for the lower spins ($I \lesim 30$).

We have here related the angular momentum dependence of the
spacings of the single-particle spectrum in the mean field, 
displayed in Fig.\ref{fig8}(b), to the behavior
of specific important orbits in the cranked potential. One may
ask whether a more consistent explanation may be achieved from
the general properties of the phase space associated with the classical
single-particle motion in the rotating potential.
So far, the questions of chaotic and regular motion in rotating deformed
potentials have only been carried out for billiards in two dimensions 
\cite{Frisk,Traiber}. It is found \cite{Frisk} 
that rotation certainly may affect the phase space generally. 
However, especially for the high kinetic 
energies of nucleon states
around the Fermi surface, the predicted effects are  small.

The absolute nature of the yrast band relative to yrare bands
of the other parity-signature configurations is worth emphasizing. We have
discussed it from our model, especially by means of Fig.\ref{fig7}.
However, it may actually be less specific to the actual
model, since it could result from a general homo-lumo gap in a 
quantal system, but now for states within an interval
of angular momenta. The favoured yrast configuration should then
be able to determine the detailed shape and other properties of 
the nucleus, and the yrare levels then have to adjust to this.

\subsection{Analysis without unfolding procedure}\label{sec:nounf}

As described in Sect.\ref{sec:form}, we apply 
the unfolding procedure in order to separate the overall excitation
energy dependence  and the local  fluctuations in the level spacings.
This procedure, however, cannot be used for the analysis
of the present experimental data in the  high spin region. In fact,
the number of  identified levels at fixed $I^\pi$ 
is far below 10 at high spins in the  experiments performed so far.
The experimental analysis in Refs. \cite{Garrett3,Garrett1,Garrett2} 
does not
apply such an unfolding procedure that is described in Sect.\ref{sec:form}.
In order  to facilitate a direct comparison between the theoretical
results and experiments, we propose
in this subsection 
another way of analysis which does not use the unfolding
procedure, but still takes into account   
the  excitation energy dependence of the level density
in an approximate way. The procedure is also applicable
to the analysis of experimental data.

Although the level density increases exponentially with
increasing  intrinsic excitation energy $U$, it may be
assumed that the level density at given $U$
is rather independent of spin and parity and nuclear species as far as
the high spin states in the same mass region are concerned.
In fact, as discussed in Ref.\cite{Level-density},
the level density at fixed $I^\pi$ can be
accounted by the level density of intrinsic configurations
in the cranking model if the spin  is sufficiently high 
(e.g. above $I \gesim 12$  for $U< 3$ MeV).
In this limit, the
level density may be approximated by the Fermi gas formula 
\cite{Level-density}
as a function of a single variable $a U$ where $a$ is the
level density parameter related to 
the single-particle level density at the Fermi surface
of the cranked mean-field.

\begin{figure}
\centerline{\psfig{figure=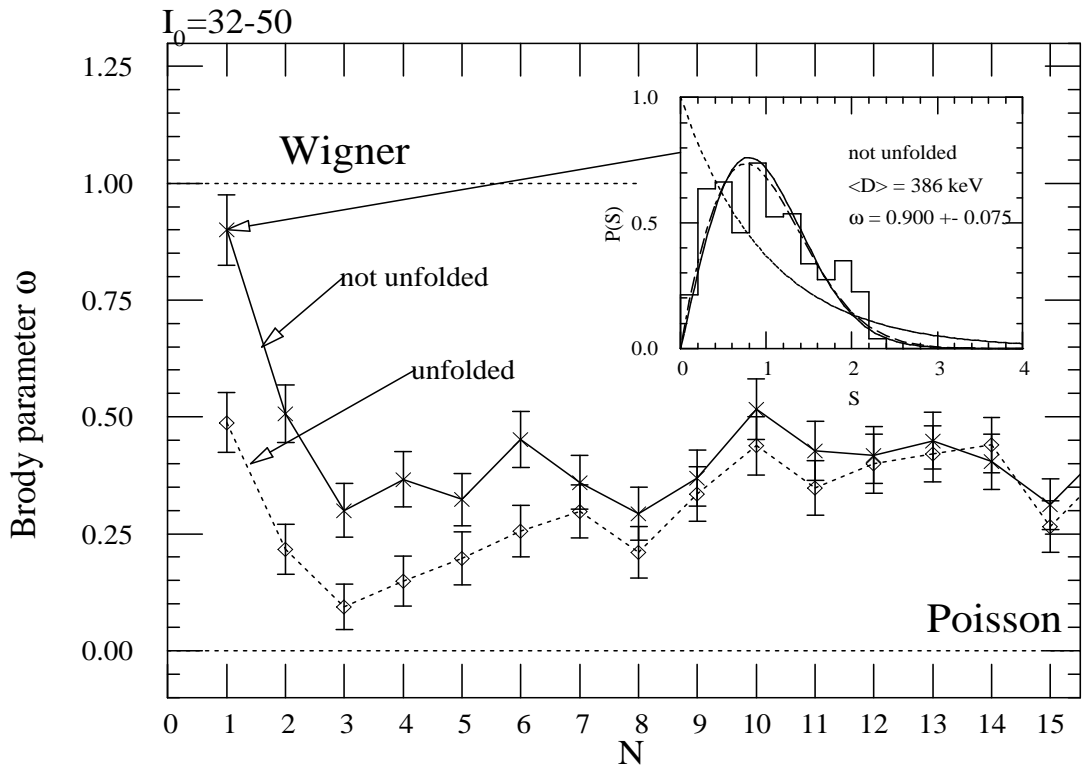,height=10cm,angle=-90}}
\caption{\label{fig9} 
The Brody parameter extracted from the NND 
obtained by using the simple normalization without
the unfolding procedure for the spin interval $I_0=32-50$ 
as a function of the
strict level ordering $N$
(the symbols connected with the solid line),
compared with that obtained with the unfolding (dotted line). 
The inset shows the NND for $N=1$ obtained without
unfolding.
}
\end{figure}

Keeping this in mind, let us consider an ensemble of
level spacings which is specified by the strict level
ordering $N$ as introduced in the previous subsections.
The level spacings within this ensemble are expected to have
a common average value since the level
ordering $N$ and hence the excitation energy is taken to be the same.
It then may be reasonable 
to define, without using the unfolding procedure,
the normalized spacing $s=D/\langle D \rangle$ by simply
dividing the spacing $D$ by
the average spacing $\langle D \rangle$ calculated for
this ensemble specified by $N$. 
We show in Fig.\ref{fig9} the NND
and the extracted Brody parameter calculated in this way.
It is seen that there exits 
small but systematic
difference for $N\lesim 5$ between the results  calculated with and
without the unfolding procedure.
The origin of the difference can be
understood by noting that the average level spacings for the
lowest states are 386, 271, 214, .. 163 keV for $N=1, 2, 3, .. 5$,
which are not very small compared with the temperature parameter
$T \sim 350 $ keV in the fitted level density. In other words, the
smooth level density $\bar{\rho}(E)$ 
varies significantly in the energy interval of the
single spacing, especially if the lowest few $N$'s are concerned. 
This causes a difference in the profile of the
NND depending on whether we adopt the unfolding
procedure or not. 
However, it should be stressed that, 
in spite of the difference depending on the way of analysis,
the Wigner-like property
associated with the yrast spacings ($N=1$) is
present in both analysis. It  becomes even more significant
with the simple way of analysis without the unfolding procedure.

\begin{figure}
\centerline{\psfig{figure=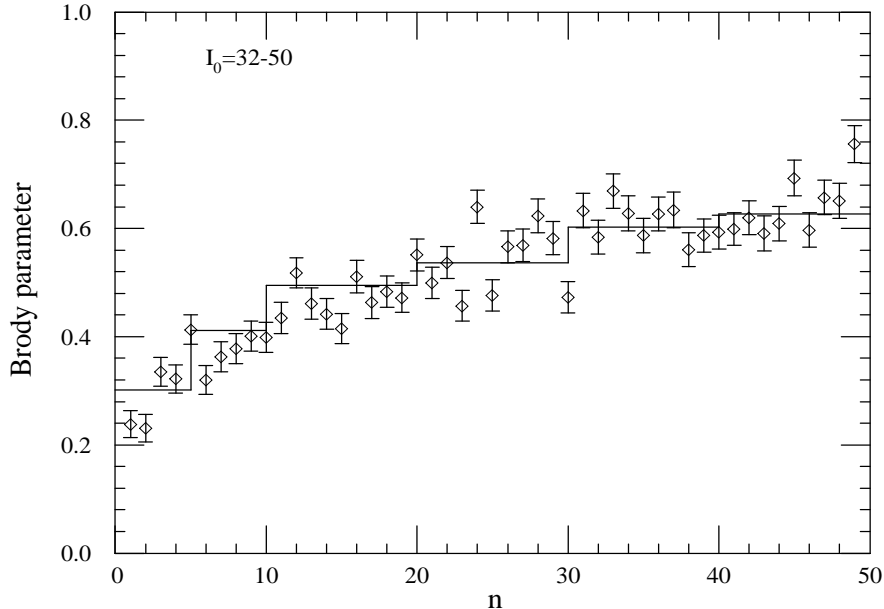,height=8cm,angle=-90}}
\caption{\label{fig10} 
The Brody parameter extracted from the
NND associated with the $n$-th level 
at each $I^\pi$ for spin interval $I_0=32-50$ (See text).
Here the unfolding procedure is not applied.  
The result  is  compared with the one with the
unfolding procedure which is calculated for the
bins of levels (same as the solid line in Fig.\ref{fig1}).
}
\end{figure}

Except for the lowest several spacings in the strict ordering, 
the NND obtained by means of the simple normalization agrees with
those obtained with the unfolding procedure.  As another illustration,
we consider an ensemble of level spacings  specified by the level
ordering $n$ defined in each spectrum for each  $I^\pi$
(note the difference between $n$ and the strict ordering $N$),
and calculate the NND and the Brody parameter 
for spin interval $I_0=32-50$ by
means of the simple normalization procedure described above.
In this case, the average spacing $\langle D \rangle$ is calculated
for each $n$. The result is compared in Fig.\ref{fig10} with
the Brody parameter (Fig.\ref{fig1} in subsect.\ref{sec:nnls})
analysed for level bins $n=1-5, 5-10, 11-20, ...$ 
by using the unfolding procedure.
Both ways of analysis agree very well with 
each other,
leading to the same conclusion about the overall excitation energy
dependence of the NND.
The NND's for $n=1$ and 2 are plotted in Fig.\ref{fig11}(a,b).

\subsection{Relation to experimental analysis}

The experimental NND obtained by Shriner et al. \cite{Shriner}
for the low-lying low-spin states 
($I \lesim 5 \hbar$)
 in rare-earth
nuclei displays a  Brody parameter around 0.3. A more recent 
analysis which includes
the observed rotational states at relatively high spins
(most of the analysed levels have $I \lesim 30$) 
reports a NND which is
close to the Poisson distribution \cite{Garrett3,Garrett1,Garrett2}.
Our theoretical calculations for higher spins
$I \gesim 30$  also  favours a  Poisson-like NND 
for the levels near the yrast line.
In the following we try to perform our theoretical analysis in a way
similar to the procedure adopted by  
Garrett et al. \cite {Garrett3,Garrett1,Garrett2}.  
One should however remark that
a comparison between our results  and the experimental findings
can only be indicative, because  they refer to different spin regions.

\begin{figure}
\centerline{\psfig{figure=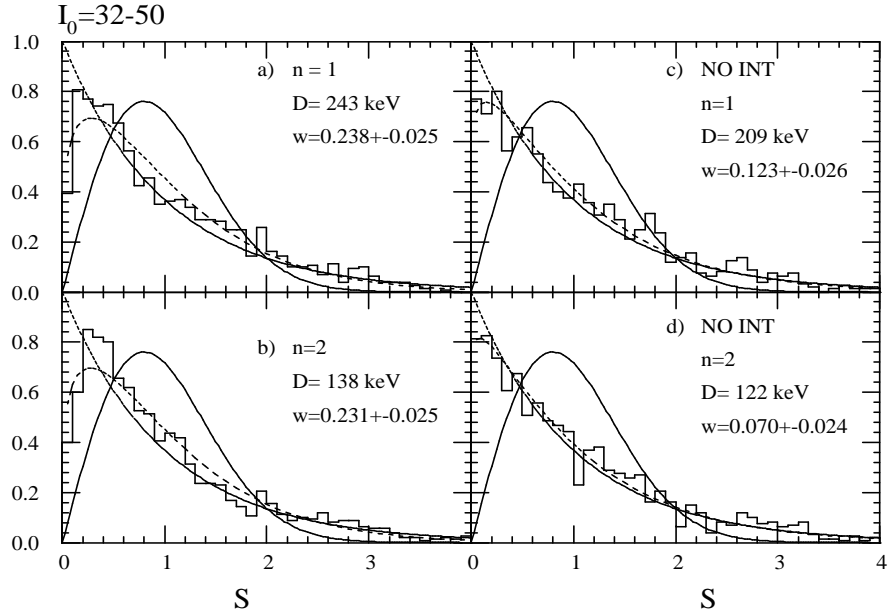,height=8cm,angle=-90}}
\caption{\label{fig11}
The Brody parameter extracted from the NND 
associated with the lowest and second lowest states ($n=1,2$)
at every $I^\pi$, for (a) and (b), respectively.
Here the unfolding procedure is not applied, and the
spacings are normalized to the average spacing defined in
each ensemble. 
(c) and (d), the same as (a) and (b) except that the
residual SDI interaction is neglected and pure mean-field
many-particle many-hole configurations are considered.
}
\end{figure}

In accordance with Ref.\cite{Garrett3}, we consider here
an ensemble of the spacings associated with the lowest and
second lowest states  at each $I^\pi$ ($n=1$ and 2). However we deal with the
spin interval $I_0=32-50$, where the pairing effects are expected 
to be weak. We remark that the pairing
effects  are removed to some extent from the experimental
analysis by excluding the lowest (0,+) spacings \cite{Garrett3}.
We adopt the normalization scheme introduced in
Subsect.\ref{sec:nounf} which do not use the unfolding procedure since
the experimental analysis \cite{Garrett3} adopt the similar
normalization.
The obtained NND's,
shown in Fig.\ref{fig11}(a,b), 
are close to the Poisson distribution, 
having Brody parameter $w \sim  0.25$
in agreement with the Poisson-like NND 
seen already for the near-yrast states.
It also shares some common features with the experimental
analysis \cite{Garrett3}: A deviation from the Poisson distribution
is seen for small spacings $s \lesim 0.2$ and is most significant for the
smallest spacing with $s\lesim 0.1$ or $D \lesim 25$ keV.

In Fig.\ref{fig11}(c,d), we also calculated the NND
in the same way except that the residual two-body interaction is
neglected. Comparing Fig.\ref{fig11}(a,b) and Fig.\ref{fig11}(c,d),
it is indicated that the deviation from the Poisson limit
at small spacings, seen in Fig.\ref{fig11}(a,b), 
is mostly caused by the residual two-body interaction.
We remark, however, that the Wigner-like distribution
associated with
the very lowest spacing $N=1$
discussed in Subsect. \ref{sec:yr} should be present, but is not visible 
neither in Fig.\ref{fig11}(a) nor in (c). 
This is because the ensemble with $n=1$ contains also the other
spacings with $N=2,3,..$, which have lower average spacing
and mask the Wigner-like distribution associated with  $N=1$ 
spacings. This suggests that, in order to find the Wigner-like
distribution caused by the mean-field effect in the experimental
analysis,  one should make an analysis by subdividing the ensemble
of the spacings with respect to the strict level order $N$.

\section{Conclusions}\label{sec:concl}

We analysed the level statistics of the high spin states
with $I \gesim 30$ in rare-earth deformed  nuclei
as a function of the intrinsic excitation energy of the
rotating nuclei.
We used a shell model approach which describes $n$p-$n$h
excitations in the cranked
Nilsson potential interacting through the surface-delta
residual interaction. 
We put emphasis on the analysis of the
near-yrast levels which may be accessible by 
discrete $\gamma$-ray spectroscopies.

The nearest neighbour level spacing distribution (NND)
and the $\Delta_3$ statistics indicate that the level
fluctuations in the near-yrast region follow a distribution
close to the Poisson limits with the extracted
 Brody parameter $w=0.2-0.3$. This value of the Brody parameter
implies significant deviation from the Poisson distribution
for spacing smaller than about $s \lesim 0.3$. 
The experimental analysis of the NND for low-spin states
\cite{Shriner} and the one including high spin rotational states
\cite{Garrett3,Garrett1,Garrett2} indicates  a 
 Poisson-like distribution
in the near-yrast states. The present analysis suggests that 
this behaviour  extends up to very high spins $I \sim 50-60$.
The level statistics approaches  the GOE limit as the intrinsic
excitation energy $U$ increases, but this process proceeds very
gradually and the chaos limit is nearly attained only with 
$U  \gesim 2$ MeV. This transition is caused by the residual
two-body interaction.

An interesting aspect of the NND emerges
when we focus on the lowest levels near the yrast line.
The level spacings between the yrast rotational band and the
next excited band with the same spin and signature  favour
a Wigner-like NND, rather than obey the
Poisson-like distribution associated with the other near-yrast levels.
The distinguishable property of the NND  associated with 
the yrast rotational bands arises from the
mean-field property of the rotating nuclei while the deviation
from the Poisson limit seen for the other spacings among
yrare rotational bands is caused by the residual two-body 
interaction. Since the lowest few levels near  yrast have
dominantly one particle excitations, the spacing
between the yrast rotational level and the next excited level
thus reflects the single-particle routhian spectrum in the
rotating mean-field at equilibrium normal deformation,
which shows a Wigner-like distribution
for the spacings between  orbits with the same parity and
signature around the Fermi surface.

\vspace{10mm}
\section*{Acknowledgments}
Discussion with J.D. Garrett is acknowledged.
One of the author, M.M., thanks the Danish Research Council
for support of his stay at Niels Bohr Institute 
where a part of the research reported here was carried out.


\vspace{10mm}
\begin{thebibliography}{99}

\bibitem{RMT}
T.A.Brody, J.Flores, J.B.French, P.A.Mello,
A.Pandy and  S.S.M.Wong, Rev. Mod. Phys. {\bf 53}(1981) 385.


\bibitem{Porter} C.E. Porter,  {\it Statistical theories of spectra:
Fluctuations} (Academic Press, 1965).


\bibitem{Mehta} M.L. Mehta, {\it Random matrices and the statistical
theory of energy levels} (Academic Press, 1967).

\bibitem{neutronres} 
R.U.Haq, A.Pandey, and O.Bohigas, Phys. Rev. Lett.
{\bf 48} (1982) 1086. 

\bibitem{billiard}
O. Bohigas, M.J. Giannoni and C. Schmit, Phys. Rev. Lett.
{\bf 52}(1984) 1. 


\bibitem{nuclchaos}
O. Bohigas and H.A. Weidenm\"uller,
Ann. Rev. Nucl. Part. Sci. {\bf 38} (1988) 421.

\bibitem{Abul} A.Y. Abul-Magd  and H.A. Weidenm\"uller, Phys. Lett.
{\bf B162} (1985) 223.

\bibitem{Shriner} J.F.Shriner,Jr., G.E.Mitchell, and T. von Egidy,
Z. Phys. {\bf A338}(1991) 309.


\bibitem{Al} J.F.Shriner,Jr., E.G.Bilpuch, P.M.Endt and
G.E.Mitchell, Z. Phys. {\bf A335}(1991) 393.

\bibitem{Sn} S. Raman, T. A. Walkiewicz, S. Kahane,
E. T. Jurney, J. Sa, Z. Gacsi, J. L. Weil, K. Allaart,
G. Bonsignori, and J. F. Shriner,Jr., Phys. Rev. {\bf C43} (1991) 521.

\bibitem{Garrett3} J.D. Garrett, J.Q. Robinson, A.J.Foglia,
and H.-Q.Jin, preprint 1996.

\bibitem{Garrett1} J.D. Garrett, J.R. German, L. Courtney, and J.M. Espino,
{\it Proc. Symp. on Future Directions in Nuclear Physics with 4$\pi$ Gamma
Detection Systems of the New Generation}, Strasbourg, 1991, eds.
J. Dudek and B. Haas (American Institute of Physics, 1992) p.345.

\bibitem{Garrett2} J.D. Garrett, 
{\it Proc. the Eighth Int. Symp. on 
Capture Gamma-Ray Spectroscopy and Related Topics},
Fribourg, 1994, ed.
J. Kern (World Scientific, 1994).


\bibitem{Bengtsson-Frauendorf} 
R. Bengtsson and  S. Frauendorf, Nucl. Phys. {\bf A314}(1979) 27;
{\bf A327}(1979) 137 .


\bibitem{Bengtsson-Ragnarsson} T.Bengtsson and I.Ragnarsson, Nucl. Phys. 
{\bf A436}(1985) 14.

\bibitem{crank-rev} As reviews, 
 S.\AA berg, H.Flocard, W.Nazarewicz,
Ann. Rev. Nucl. Part. Sci, {\bf 40} (1990) 439; \\
Z.Szymanski, {\it Fast Nuclear Rotation} (Clarendon Press, 
Oxford, 1983); \\
M.J.A. de-Voigt, J.Dudek, Z.Szymansky, Rev. Mod. Phys.
{\bf 55} (1983) 949.




\bibitem{Lauritzen} B.Lauritzen, T.D\o ssing and R.A.Broglia, Nucl. Phys. 
{\bf A457}(1986) 61.


\bibitem{FAM} B.Herskind, A.Bracco, R.A.Broglia, T.D\o ssing,
A.Ikeda, S.Leoni, J.Lisle, 
M.Matsuo, and E.Vigezzi, Phys. Rev. Lett.
{\bf 68}(1992) 3008; \\
T. D{\o}ssing, B. Herskind, S. Leoni, M. Matsuo,
A. Bracco, R.A. Broglia, and Vigezzi,
Phys. Report {\bf 268} (1996) 1.

\bibitem{Aberg} S.\AA berg, Phys. Rev. Lett. {\bf 64}(1990) 3119;\\
 S.\AA berg, Prog. Part. Nucl. Phys. vol.28 (Pergamon 1992) p.11.

\bibitem{Matsuo96} M. Matsuo, T. D{\o}ssing, E. Vigezzi, R.A. Broglia,
        and K. Yoshida, preprint YITP-96-48 1996, Nucl. Phys. in press.


\bibitem{Mozkowski} I. M. Green and S. A. Mozkowski, Phys. Rev. {\bf 139}
(1965)B790; \\
R.Arvieu and S.A.Mozkowski,Phys. Rev. {\bf 145} (1966) 830.

\bibitem{Faessler} A. Faessler, Fortschr. Phys. {\bf 16}(1968) 309.

\bibitem{Matsuo93} M. Matsuo, T. D{\o}ssing, E. Vigezzi and R.A. Broglia,
                    Phys. Rev. Lett. {\bf 70} (1993) 2694; \\
M. Matsuo, T. D{\o}ssing, B. Herskind, 
S. Frauendorf,  E.  Vigezzi  and  R.A.  Broglia,
 Nucl. Phys. {\bf A 557} (1993) 211c.


\bibitem{Bracco}
 A. Bracco,  P. Bosetti, S. Frattini, E.~Vigezzi,
S.~Leoni, T.~D\o ssing,
B.~Herskind, and  M.~Matsuo, Phys. Rev. Lett. {\bf 76} (1996) 4484.


\bibitem{ibm} T.Mizusaki, T.Otsuka, and P.von Brentano,
Nucl. Phys. {\bf A598} (1996) 47.

\bibitem{IBFM} A.Alhassid and D.Vretenar, Phys. Rev. 
{\bf C46} (1992) 1334.

\bibitem{Kruppa} A.T.Kruppa, K.F.Pal, N.Rowley, Phys. Rev. 
{\bf C52} (1995)1818.


\bibitem{Garrett-pair} J.D.Garrett, {\it Nuclear Structure 1985},
eds. R. Broglia, G. B. Gagemann and B. Herskind
(Elsevier Science,1985) p.111.

\bibitem{Shimizu} Y.R.Shimizu, J.D.Garrett, R.A.Broglia, M.Gallardo and
E.Vigezzi, Rev. Mod. Phys. {\bf 61} (1989) 131.

\bibitem{Shimizu-Oak} Y.R.Shimizu, Nucl.Phys., {\bf A520} (1990) 477c.



\bibitem{PES} S. \AA berg, Phys. Scr. {\bf 25} (1982) 23.

\bibitem{Werner} T.R. Werner and J. Dudek,
Atomic Data and Nuclear Data Tables, {\bf 50}(1992) 179.


\bibitem{Def-parm} R.Bengtsson, S.Frauendorf and F.-R.May,
Atomic Data and Nuclear Data Tables, {\bf 35}(1986) 15.

\bibitem{unfolding}
O. Bohigas and  M.J. Giannoni, {\it Mathematical and Computational
Methods in Nuclear Physics, Lecture Notes in Physics 209},
eds. J.S.Dehesa, J.M.G. Gomez and A.Polls (Spriger 1984)  p.1.

\bibitem{CTF}
 A.Gilbert and A.G.W.Cameron, Can. J. Phys. {\bf 43} (1965) 1446.


\bibitem{delta3} F.J. Dyson and M.L. Mehta, J. Math. Phys. {\bf 4}
(1963) 701.

\bibitem{Pe95} P.Persson and S. \AA berg, Phys. Rev {\bf E52} (1995)
148.

\bibitem{Berry} M.V. Berry, Proc. R. Soc. Lond., {\bf A400}(1985) 229.

\bibitem{Arve} P. Arve, Phys. Rev. {\bf A44}(1991) 6920. 


\bibitem{Fitz168Yb} A.~Fitzpatrick, S.A.~Araddad, R.~Chapman, J.~Copnell,
F.~Lind{'{e}}n, J.C.~Lisle, A.G.~Smith, J.P.~Sweeney, D.M.~Thompson,
W.~Urban and S.J.~Warburton, J.~Simpson, C.W.~Beausang, J.F.~Sharpey-Shafer,
S.J.~Freeman, S.~Leoni, and J.~Wrzesinski,\\
Nucl. Phys. {\bf A585} (1995)335.

\bibitem{Nord164Yb} A. Nordlund, R. Bengtsson, P. Ekstr{\"{o}}m,
M. Bergstr{\"{o}}m, A. Brockstedt, H. Carlsson, H. Ryde, Y. Sun, A. Atac,
G.B. Hagemann, B. Herskind, H.J. Jensen, J. Jongman, S. Leoni,
A. Maj, J. Nyberg and P.O. Tj{\o}m,
Nucl. Phys. {\bf A591}(1995)117.


\bibitem{Frisk} H. Frisk and R. Arvieu, J. Phys. {\bf A22} (1989)
1765;  \\
H. Frisk and R. Arvieu, Nucl.  Phys. {\bf A495} (1989) 291c.

\bibitem{Traiber} A.J.S. Traiber, A.J. Fenrdik, and
M. Bernath, J. Phys. {\bf A23} (1990) L305.

\bibitem{Level-density} S.\AA berg, Nucl. Phys. {\bf 477} (1988) 18.



\end{thebibliography}
\end{document}